\documentclass{article}
\usepackage{latexsym}
\usepackage{amssymb}
\usepackage{graphics}
\newcommand{\be}{\begin{equation}}
\newcommand{\ee}{\end{equation}}
\newcommand{\bea}{\begin{eqnarray}}
\newcommand{\eea}{\end{eqnarray}}
\newcommand{\cg}{\mathcal{G}}
\title{Noise Kernel for Reissner Nordstr\"{o}m Metric: Results at Cauchy 
Horizon}
\author{Seema Satin \\
\tiny{sesatin@tku.edu.tw}\\
\small{ Dept. of Physics, Tamkang University, Tamsui Distict, Taipei,
 Taiwan (ROC)
}
}
\date{ } 
\begin{document}
\maketitle
\begin{abstract}
We obtain  point separated Noise Kernel for the Reissner Nordstr\"{o}m metric.
 The Noise Kernel defines the fluctuations of the quantum stress tensor
and is of central importance to Semiclassical Stochastic Gravity.
 The metric is  modeled as gravitationally collapsing
 spacetime, by using suitable coordinate transformations, defined earlier.
 The fluctuations of the quantum stress tensor, at the  final stage of collapse
 are then analysed for both, the naked singularity and black hole end states.
 The behavior of this Noise Kernel, at the Cauchy Horizon for naked singularity
 shows markedly different behaviour from  self similar Tolman Bondi metric,
which was obtained earlier. In
 the latter
  a very unique divergence was seen, which  does not appear for the
Reissner Nordstr\"{o}m metric, here .  It  is known that the
quantum stress tensor itself, diverges at the Cauchy Horizon (CH) for both
 of these metrics  . In contrast, it can now be seen that the
 the fluctuations of the stress tensor behave differently for
 the two cases.
 We give a discussion and further directions for investigations of 
 this interesting behaviour in the two cases (regarding the collapse
 scenario).
\end{abstract}
\section{Introduction}
The Cosmic Censorship Conjecture (CCC) is considered to be an important open
 area
of research since its inception in 1969. The precise mathematical 
formulation of this hypothesis is much awaited . The two possible end states
for complete graviational collapse of a massive star, namely
 black hole (BH) and naked singularity (NS), need an elaborate and varied
 analysis for the confirmation  of the CCC or otherwise.
There are several studies in this direction both numerical and analytical, 
 which
 span classical and quantum domains \cite{shap,new,Ori,Ori2,Pankaj}. What 
prevents naked singularities to occur in nature, according to the CCC is a
 quest. 

However, it is important to note that, the naked singularities arise not only
as solutions to the Einstein's equations for a collapsing star, 
but are also known to have observational consequences \cite{vir1,vir2}.
 
The genericity and stability criteria for naked vs covered singularities 
 is a vast literature \cite{pankaj2,ritu,joshi,suneeta} nevertheless,
 some new  insights can shed light on this study from different perspectives. 

After the formulation of Semiclassical Stochastic Gravity \cite{bei1} in late 
90's, the  current focus now ,is over the  applications of the theory to
 Cosmology and black hole (BH) physics.
 The Noise Kernel, defines the fluctuations of the quantum stress tensor in a 
given spacetime, and is of central importance to Stochastic Gravity 
\cite{Bei2}. In this mean field approach towards gravity, the 
background spacetime plays the role of the system, while the quantum  states
that of the environment. 

Stochastic Gravity is very new to the study of CCC, and has 
recently  been seen to contribute towards interesting directions in this 
regard .
The first of its applications to CCC has been carried out in  
 \cite{kinjalk}, for the  spherically symmetric dust (self-similar Tolman
 Bondi metric).
The results therein show a marked difference in the behaviour of 
fluctuations of the quantum stress tensor, for the two possible end states.  
One sees a unique divergence at the CH for the Naked Singularity (NS) while
no such divergence is seen if the collapse results in a black hole.  
This result awaits explanation for such a behaviour at CH, where the validity
 of stochastic semiclassical gravity breaks down. 

In this article, we attempt to find out if a similar divergence at the CH, 
 occurs in case of Reissner Nordstr\"{o}m metric also.  This would further 
help investigations towards the earlier obtained result and indicate if such 
a divergence is a generic feature of naked singularities.
 
One would expect such a  divergence of the Noise Kernel at the CH  for all 
cases, should that be a peculiar behaviour for all NS end states.
Incidentally, we obtain a very different result in the case of 
 Reissner Nordstr\"{o}m metric. No such divergence of the Noise Kernel is seen
 here. This raises a lot of further interest, into underlying reasons for such
 behaviour. Different spacetimes metrics can be associated with complete 
gravitational collapse that admit NS and BH solutions as end states.

This study gives directions for more to be explored, and a theme which 
looks into specific physical conditions responsible for the kind of difference
 that we see. 
Possible reasons for the difference in the behavior of fluctuations of stress
tensor  at the CH, for the two spacetimes (namely Tolman Bondi and RN metric) 
 are given towards the end of the article . 

The article is organised as follows.

In section (\ref{sec:metric}) we present the Reissner Nordstr\"{o}m metric
 solution in a form, more suitable for the collapse scenario. In section
 (\ref{sec:NK})
a short desciption of the Noise Kernel, defining the fluctuations of stress
 tensor is given. In the subsequent section (\ref{sec:QS}) the relevant
 quantum states associated with the stress tensor and related issues are
 addressed. The metric is obtained in appropriate form, which is convenient for
 the  calculation  technique used  for  Noise Kernel.  
Section (\ref{sec:inter}) gives the physical interpretation of the result
 obtained at the CH. In the last section, we give the conclusions that can be
 drawn from the result obtained and discuss some important issues. 
\section{Evaluating the Noise Kernel for Reissner Nordstr\"{o}m metric}
In the following, we attempt to calculate the Noise Kernel components
for the Reissner Nordstr\"{o}m metric in detail. The     
procedure for this  and relevant issues are reviewed. As
 mentioned  later, a symbolic code has been used to evaluate the final
 expressions.
\subsection{The Reissner Nordstr\"{o}m metric and gravitational collapse} 
\label{sec:metric}
The Reissner Nordstr\"{o}m (RN) solution in coordinates suitable for dynamical
 evolution of the  initial data is developed in \cite{prasanna}. This is
necessary for the gravitational collapse scenario applied to the spacetime
metric.  We briefly review these coordinates and the metric stucture in the
 following.

The Reissner Nordstr\"{o}m solution can be treated as the evolution
 of a spherically symmetric metric with charge $E$ and mass $M$ from a 
regular initial data for $ E<M $ . This metric can be written
in coordinates which mimic dynamical evolution. The background metric, along
 with  suitable quantum states, descibes the semiclassical behaviour of such an 
evolution.  
  This can be viewed as resulting from initial data, evolving as a collapse
 leading to curvature singularity. 

The RN metric is usually represented as
\be
ds^2 = P(r) dt^2 - P(r)^{-1} dr^2 - r^2 d\Omega^2
\ee
where $P(r) = 1- 2M/r + E^2/r^2 $. As introduced in \cite{prasanna} the special coordinates defined by  
\bea 
T &=& t+ \int \frac{g(r)}{P(r)} dr \\
R &=& t+ \int \frac{1}{g(r) P(r)} dr
\eea
are uselful in casting the metric as initial value problem. The function 
$g(r)$ has been introduced and engineered  so as to ensure
regularity of the transformations proposed in the above reference. It is chosen
in such a way so that $P(r)/(1-g^2(r)) > 0$, this maintains the signature of the
transformed metric.

The metric thus takes the form.  
\be
ds^2 = \frac{P(r)}{1- g(r)^2} dT^2 - \frac{g(r)^2 P(r)}{1- g(r)^2} dR^2 - r^2 
d\Omega^2
\ee
Since we aim at analysing the same physical situation of the gravitational
 collapse as in \cite{prasanna}, we  intend to use the same coordinates.

Further we use the double null coordinates ,
\be 
u= T-g(r) R, v= T+g(r)R
\ee
 to obtain the form  
\be
ds^2 = \frac{P(r)}{1-g(r)^2} du dv + r^2 d\Omega
\ee
for the metric.

These coordinates have been used to analyse the behavior of quantum stress 
tensor near the cauchy horizon. It is seen in the above reference that the
stress tensor diverges at the CH. It remains regular at the event horizon
when a covered singularity is formed. Though the CH is unstable here, the study
of fluctuations is of interest because it probes the extended spacetime 
structure in the neighbourhood of the CH, as explained later . 
The fluctuations of the stress tensor, which are responsible for the induced
 fluctuations of the metric, can be characterised by the Noise Kernel in the
 Einstein Langevin formalism. 

(A similar study for the  Tolman Bondi metric shows that 
 the quantum stress tensor  and noise kernel \cite{suktb,kinjalk} both diverge
at the CH  . 
 
For regions near the CH, $g=g_1$ (detials as in \cite{prasanna}) in the above
 equation and so we can
write
\be \label{eq:4}
ds^2 = \frac{P(r)}{1-g_1^2} du dv + r^2 d\Omega
\ee

Next, we give a brief description of the Noise Kernel bitensor, which forms a
quantity of central interest in our work and Stochastic gravity. 

\subsection{The Noise Kernel} \label{sec:NK}
The Einstein Langevin Equation defining Semiclassical stochastic gravity, is an 
extension to the theory of semiclassical gravity. The inclusion of fluctuations
 of 
the quantum stress tensor, gives semiclassical theory a Langevin
approach. These induced fluctuations to the  metric, play an imporant role
in determining extended structure of the spacetime. The Einstein Langevin
equation  reads
\be
G_{ab} + \Lambda g_{ab} = \langle \hat{T}_{ab} \rangle +  \hat{\xi}_{ab}
\ee
(we use the convention $ c=G=\hbar = k_B =1 $)
where expectation is taken over a normalized state and $\hat{\xi}_{ab}$ is a
 random variable.
This is characterised by
\be 
\langle \hat{\xi}_{ab}(x) \rangle_s = 0 , \langle \hat{\xi}_{ab}(x) \hat{\xi}_{cd}(x') \rangle_s = N_{abc'd'} (x,x')
\ee
The source in the above, thus is gaussian and the bitensor $ N_{abc'd'}(x,x') $
expression can be obtained explicitly as follows
\be
N_{abc'd'} (x,x') = \frac{1}{2} \langle \{ \hat{t}_{ab}(x), \hat{t}_{c'd'}
(x') \} \rangle
\ee
where 
\[ \hat{t}_{ab} (x) \equiv \hat{T}_{ab}(x) - \langle \hat{T}_{ab} (x) \rangle \]
The general expressions for the Noise Kernel in non-coincident limit have been
 obtained in \cite{Bei2}, giving various cases of of couplings, for
massive as well as massless fields .

Here we are interested in calculating the Noise Kernel for conformally
 invariant scalar field.   

To begin with, the classical stress tensor of a conformally invariant scalar
 field $\phi$ is given by
\be
T_{ab} = \nabla_a \phi \nabla_b \phi - \frac{1}{2} g_{ab} \nabla^c \phi
\nabla_c \phi + \frac{1}{6}( g_{ab} \Box - \nabla_a \nabla_b + G_{ab} ) \phi^2
\ee
The scalar field after being quantised is treated as an operator, while
the metric $g_{ab}$ is treated classically.

The quantum states need to be specified for conformally invariant fields.
The Noise Kernel can then be obtained in terms of Wightman function 
\cite{Bei2}, and the expression is given as  
\be \label{eq:NK}
N_{abc'd'} = Re\{ \bar{K}_{abc'd'} + g_{ab} \bar{K}_{c'd'} + g_{c'd'}
\bar{K}'_{ab} + g_{ab} g_{c'd'} \bar{K} \}
\ee
\bea 
9 \bar{K}_{abc'd'} & = & 4 (\cg_{;c'b} \cg_{;d'a} + \cg_{c'a} \cg_{;d'b}) 
+ \cg_{;c'd'} \cg_{;ab} + \cg \cg_{;abc'd'} - \nonumber \\ 
& &  2( \cg_{;b} \cg_{c'ad'} + \cg_{;a} \cg_{;c'bd'} + \cg_{;d'} \cg_{;abc'} +
 \cg_{;c'} \cg_{;abd'}) + 2( \cg_{;a} \cg_{;b} R_{c'd'} + \nonumber \\
 & & \cg_{;c'} \cg_{;d'} R_{ab})- (\cg_{;ab} R_{c'd'} + \cg_{;c'd'} R_{ab} )
 \cg + \frac{1}{2} R_{c'd'} R_{ab} G^2  \label{eq:NK1}
\\
36 \bar{K}'_{ab} &=& 8 ( -\cg_{;p'b} \cg^{p'}_{; a} + \cg_{;b} \cg_{;p'a}^{p'}
+ \cg_{;a} \cg_{;p'b}^{p'} ) + \nonumber \\
& & 4 ( \cg_{;}^{p'} \cg_{;abp'} - \cg_{;p'}^{p'} \cg_{;ab} -
 \cg \cg_{;abp'}^{p'} ) - 2 R' (2 \cg_{;a} \cg{;b} - \cg \cg_{;ab}) \nonumber\\
& & -2 ( \cg_{;p'} \cg-;^{p'} - 2 \cg \cg_{;p'}^{p'} ) R_{ab} -
 R' R_{ab} \cg^2  \label{eq:NK2} \\
36 \bar{K} & = & 2 \cg_{;p'q}\cg_;^{p'q} + 4 ( \cg_{;p'}^{p'} \cg_{;q}^q +
\cg \cg_{;p q'}^{p q'}) - 4 ( \cg_{;p} \cg_{;q'}^{pq'} + \cg_;^{p'}
 \cg_{;q p'}^q ) \nonumber \\
& &  + R \cg_{;p'} \cg_;^{p'} + R' \cg_{;p}
 \cg^{;p} - 2(R \cg_{;p'}^{p'} + R' \cg_{;p}^p) \cg + \frac{1}{2} R R' 
\cg^2  \label{eq:NK3}	
\eea
Here the  Noise Kernel obeys the following properties \cite{efte}. 
\begin{enumerate}
\item $ N_{abc'd'} (x,x') = N_{c'd'ab}(x',x)$
\item $ \nabla^a N_{abc'd'} =  \nabla^b N_{abc'd'} = \nabla^{c'} N_{abc'd'}
= \nabla^{d'} N_{abc'd'} =0 $.
\item $ N^{a}_{\hspace{.2cm}ac'd'} =
 N_{ab\hspace{.2cm}c' }^{\hspace{.3cm}c'}=0 $.
\item  The Noise Kernel is semidefinite ,

$ \int d^4x \sqrt{-g(x)} \int d^4x' \sqrt{-g(x')} f^{ab}(x)
 N_{abc'd'}(x,x')$
$ f^{c'd'}(x'x) \geq 0 $  

  for any real tensor field $f^{ab}(x)$. 
\end{enumerate} 
\subsection{The Quantum states and  Wightman functions} \label{sec:QS}
The quantum states play a decisive role in the behavior of the fluctuations
of the matter fields . In our study, we try to use the most 
general class of  quantum states for non-interacting fields. These fulfill the
 basic requirements
  for a well defined stress tensor and  its fluctuations, in the 
curved spacetime background. We therefore use, the Quasifree thermal states of
 Hadamard type in this work.
The states being Hadamard ensures that the stress tensor is well defined
within maximal Cauchy development and obeys Wald axioms.

The Wightman two point function  in (\ref{eq:NK1}) - (\ref{eq:NK3}) is given by
\be
\cg (x,x') = \langle \phi(x) \phi(x') \rangle
\ee
this determines the quantum state of the field if we work with quasi-free
(Gaussian) states. 

The state being thermal or KMS type assigns a temperature $\kappa$. 
This leads to  Wightman functions, being approximated for thermal states via
the Gaussian approximation (Page's approximation) as described in \cite{efte}.
The temperature is given by
\be
T= \frac{\kappa}{2 \pi}
\ee
An appropriate form for Wightman function for given  spacetime is
required to obtain the noise kernel. In general, numerical methods can be
used to get this for arbitrary separtions. For small separations however, 
analytical form for the expressions can be obtained by using an 
 approximate method. This has been  established  in earlier  work, 
and we intend to use  Page's
 approximation as has been done for  other spacetimes \cite{kinjalk,efte}.

The  Wightman function  for ultrastatic spacetime, which can be related
 conformally to the Reissner Nordstr\"{o}m metric is calculated. This
is then used to obtain the Noise Kernel for the same.  
These Wightman functions are, as shown below, expressed in terms of the synge 
function.
However, first we need to put the metric in the desired form as would be used.

For this, we  transform the Reissner Norstdrom metric to ultrastatic form.
  Following equation (\ref{eq:4})
\be
ds^2 = P(r) du_+ dv + r^2 d\Omega
\ee
where $du_+ = du/(1-g_1^2) $,

let $P(r)/r^2 du_+ = dU $, then
\be
ds^2 = r^2[dU dv + d\Omega] 
\ee 
Further transforming as,

 $T= U-v$ and $X= U+v$ ,
\[ U = \frac{T+X}{2} , v = \frac{X-T}{2} \]
which gives
\[ dU dv = \frac{1}{4} (-dT^2 + dX^2 ) \]
the form of the metric is now
\be \label{eq:ultra}
ds^2 = r^2 \{ -\frac{1}{4} dT^2 + \frac{1}{4} dX^2 + d\Omega \}
\ee

This is in conformal ultrastatic form,
\be
ds^2 = \Sigma^2 \{ -\frac{1}{4} dT^2 + \frac{1}{4} dX^2 + d\Omega \}
\ee
where the conformal factor $\Sigma = r^2$. We compare this with  the
 ultrastatic  metric which takes the form
\be
ds^2 = dt^2 + g_{ij}(\vec{x}) d\vec{x}^i d\vec{x}^j 
\ee 
where metric functions $g_{ij}$ are independent of time $t$. For this metric
the synge function takes the form
\be
\sigma(x,x') = \frac{1}{2} ((t-t')^2 - \bf{r}^2)
\ee
where $\bf{r}^2$ is twice the spatial part of the Synge function, which
depends only on the spatial coordinates.

The expression for Wightman function in ultrastatic background metric 
for a KMS state can be calculated under Gaussian approximation \cite{efte}
\be
\cg(\Delta t,x,x') =  \frac{\kappa \sinh(\kappa \mathbf{r})}{8 \pi \mathbf{r}
[\cosh(\kappa \mathbf{r})- \cosh(\kappa \Delta t)] } U(\Delta t, x,x')  
\ee
This can be expanded as 
\be \label{eq:wight}
\cg(x,x') = \frac{1}{8 \pi^2} [ \frac{1}{\sigma} + \frac{\kappa^2}{6} -
\frac{\kappa^4}{180} ( 2 (\Delta t)^2 + \sigma) + O[(x-x')^4]] U(x,x')
\ee
where 
\bea \label{eq:U1} 
U(x,x') &=& \Delta^{1/2} (x,x')  \nonumber \\ 
\Delta(x,x') &=& \frac{1}{\sqrt{-g(x)} \sqrt{-g(x')}} \det(\sigma_{;ab'})
\eea
The expressions as presented above, are made up of the synge function.
For the above metric (\ref{eq:ultra}) in ultrastatic form , 
 we calculate this synge function and it takes the form

\be \label{eq:synge}
\sigma = - \frac{1}{2} (T'-T)^2 + (X'-X)^2 + \eta^2
\ee 
where 
\[ \cos(\eta) = \cos(\theta) \cos(\theta') + \sin(\theta) \sin(\theta')
\cos(\phi- \phi') \]
Given this function as in (\ref{eq:synge}), we substitute it in equation
(\ref{eq:U1}) and  get
\be \label{eq:U}
U(x,x') = \sqrt{\frac{\eta}{\sin(\eta)}}
\ee
The Noise Kernel components for the RN metric can now be computed using the
 expressions above.
\subsection{Noise Kernel for Reissner Nordstr\"{o}m metric}
Substituting equation (\ref{eq:synge}) and (\ref{eq:U}) in (\ref{eq:wight}), and
finally the expression for Wightman function thus obtained in (\ref{eq:NK1})
(\ref{eq:NK2}), (\ref{eq:NK3}) and (\ref{eq:NK}) different components of 
the Noise Kernel can be obtained.

We display here, one of the components thus evaluated. Other components are
similar in nature, as far as the results at the CH discussed in
 this paper are concerned. The following component $N_{TTT'T'}$ which 
we show below, is point separated with $\eta = 0 $ and 
$\delta X = 0$, while $\delta T \neq 0$. The result below is obtained
after conformal transformation of Noise Kernel from ultrastatic to the original
spacetime, 
\be 
N_{abcd}(x,x') = \Sigma^{-2}(x) \tilde{N}_{abcd}(x,x') \Sigma^{-2}(x')
\ee
\bea 
N_{TTT'T'} &=& \frac{1}{r^2(T,X) r^2(T',X)}[ \kappa^0(\frac{2897}{288 \pi^2 \delta T^8}-\frac{1}{72 \pi^2
 \delta T^7}+\frac{349}{1152 \pi^2 \delta T^6}-\frac{7}{3456 \pi^2 \delta T^5}
\nonumber \\
& & +\frac{116640000+13934592000 \pi }{15925248000 \pi^2 \delta T^4})
\nonumber \\
 & & \kappa^2 (\frac{1}{96 \pi ^2 \delta T^8}+\frac{211}{864 \pi ^2 \delta T^6}
-\frac{1}{864 \pi ^2 \delta T^5}+\frac{7}{512 \pi ^2 \delta T^4}) \nonumber \\
& & \kappa^4 (1/(1152 \pi^2 \delta T^6) + 1133/(34560 \pi^2 \delta T^4) )
\nonumber \\
& &+ \kappa^6 ( 1/(34560 \pi^2 \delta T^4) )]
\eea 
These expressions  have been evaluated  using symbolic code developed for the
same. One can check  correctness of the result by using properties of the
 Noise Kernel given in the earlier section.
\section{Interpretation of the Noise Kernel expression at the Cauchy Horizon}
\label{sec:inter}
\begin{figure}
\resizebox{\columnwidth}{!}{\includegraphics{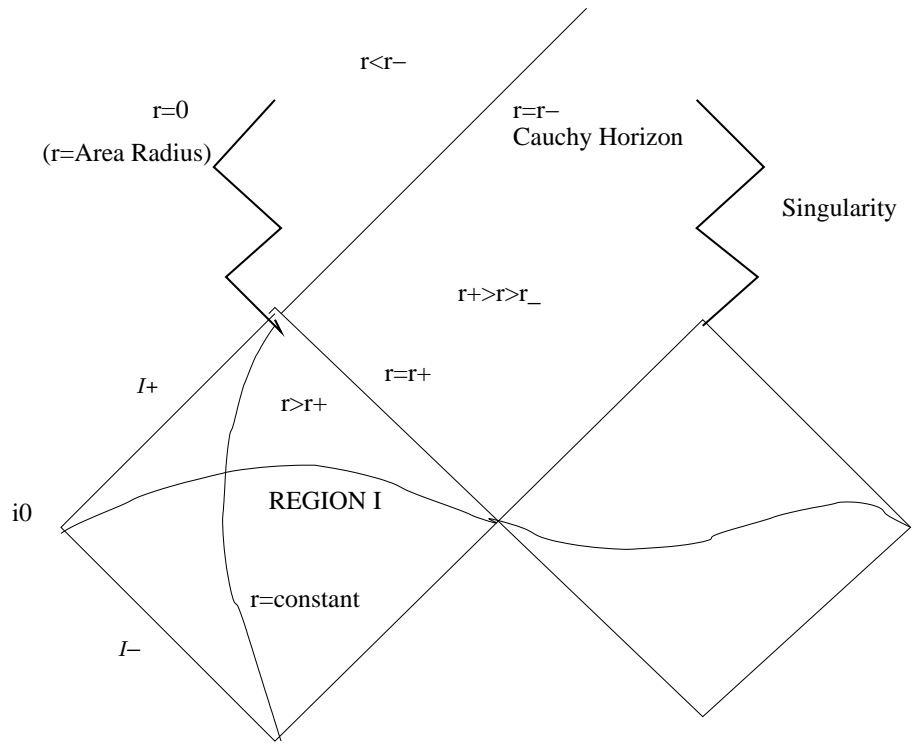}}
\caption{Penrose diagram of the Reissner Nordstr\"{o}m Metric  for the case
$E<M$. Cauchy surface across region I is shown by the curved line. The  CH is
shown by $r=r_-$.
}
\label{fig:RN}
\end{figure}
The equation for the Cauchy Horizon is given by $P(r) =0$, where the smaller
root 
 $ r= r_- $ as shown in the figure defines the first null ray (i.e the CH)
coming out of the singularity \cite{Haw}. If we take $r(T,X) = r_-$, 
placing one point on the CH and the other at $r(T',X)$ separated from it
( as valid under gaussian approximation)  we 
see that the Noise Kernel is regular everywhere including the CH. This 
result is 
markedly different from that of the Tolman Bondi metric.  We ensure
that all components of the point separated Noise Kernel for Reissner 
Nordstr\"{o}m metric are regular at the CH and no divergences are seen to occur.

The importance of the above result lies in realising that, though the Cauchy
Horizon here is unstable, we can draw some very important conclusions on
the regions of spacetime  very near to  the naked singularity.  Any other
 analysis fails or is invalid in this region. The noise kernel is the 
correlation of the fluctuations
at two different points. The above calculation is valid only for spacelike
or time like separations. This ensures that only one point can be put on the
 CH,
and the other should be placed on a timelike or a spacelike separation nearby.
We see that this correlation is finite , even though the CH itslef is unstable
 with the quantum stress tensor also being divergent. The result thus obtained
 indicates that, using the Einstein Langevin approach it is still possible to
 probe the extended structure of spacetime in terms of backreaction studies.
More importantly, this is possible at the stochastic semiclassical level,
 where qunatum gravity effects set in .
 (This is certainly not possible for
 the Tolman Bondi metric, since the noise kernel is divergent there).  

On the other hand, noise kernel divergence (as for Tolman Bondi metric) 
indicates large contributions of fluctuations near CH.
 The question which would naturally arise is regarding the nature of
 probability distribution of the stress tensor (which is treated as a random 
variable).
In 3+1-dimensions, a closed form expression for the probability distribution
 for the stress tensor is not available, but a certain amount is known
about the behaviour of the moments of the energy density e.g., for massless 
scalar and electromagnetic fields. This implies that the probability
 distribution has a slowly decaying tail at large positive values \cite{ford}.
 Ford argued from this, that large positive fluctuations would  dominate over
 thermal fluctuations at large energies, with potential implications for rate
 estimates of black hole nucleation from the vacuum.

The noise kernel for a thermal state becoming very large  would imply  that
 thermal fluctuations seem to dominate near CH. This result is complementary to Ford's argument above. 

  The question that one could ask here is,  whether the noise kernel blows up
 at every Cauchy horizon. This paper shows that the answer is negative. The
 Reissner Nordstr\"{o}m metric is an exception.

\section{Conclusion and further Directions}

The CH of the RN metric is important in some other ways as well. The classical
 phenomenon of mass inflation of the RN singularity has been well known. The 
large blue shift of inflalling radiation leads to such a phenomenon. Physically, particle
creation must also be addressed, especially when the blue shift energies are
 very high. Quantum stress tensor is a way to address both blue shift energy
 as well as particle creation effects together . 
   But the quantum stress tensor also diverges \cite{prasanna} upholding the 
physical implication that the CH will behave like a singular light like 
surface (what Penrose terms as a "thunderbolt"). 
    In this analysis, quantum fluctuations had not been addressed. Like dust
 collapse where divergent contributions of fluctuations put in doubt the
 average quantum stress results, we are compelled to ask if fluctuations call
 into question  the idea of thunderbolts or mass inflation. Our results show 
that they do not.

We have obtained the expressions for Noise Kernel in case of Riessner 
Nordstr\"{o}m
metric while it is been  modeled for complete gravitational collapse. The Noise
 Kernel is regular everywhere in the spacetime and thus backreaction
studies, using the Einstein Langevin equation can be 
carried out here. It would be interesting to see the backreaction effects of
 induced fluctuations of the stress tensor at the Cauchy Horizon . This is a 
very important result, for various reasons . 
\begin{itemize}
\item The quantum stress tensor for the RN metric  diverges (ref. to earlier
 work) at the CH, while the fluctuations of the same do not. This indicates 
that the blowing up of the quantum stress tensor can be interpreted as energy
burst more appropriately in the case of RN metric as has been suggested. While 
for the Tolman Bondi metric the divergence of the stress tensor as well as its
 fluctuations do not give  clear indication for the energy burst at the CH.
\item 
 One may
observe that, quantum fluctuations diverge at CH, in the case where classical 
stress tensor is non zero. While it is not so  
when the classical stress tensor is zero. In order to confirm this 
behaviour, we suggest few more examples of spacetimes to be worked
out on similar lines.
Further investigations  can be carried out based on the results thus obtained.
 We intend to explore in this direction in future. 
\item It is important to note that this distinction in the gravitational
collapse  for self similar Tolman Bondi and Reissner Nordstr\"{o}m metric is
 seen
only for the "fluctuations" of the quantum stress tensor, while the quantum
stress tensor itself is divergent in both cases (as obtained in  earlier work).
\item Corresponding to the collapse scenario,  difference between the two 
metrics compared here,  occurs
first at the classical level, while remaining same at the
semiclassical level and then reappearing at the stochastic semiclassical level.
It is important to ponder over the reason and mechanism, how the classical
 and the quantum matter fields are connected together to the
 spacetime geometry and if they influence each other via the background
 spacetime. 
This may lead one to investigate the basic structure of spacetime geometry and
related matter fields in the non-linear Einstein equation at the classical
, semiclassical and stochastic level. Based on this it may further be 
interesting to address fundamental issues relating these three domains. 
\end{itemize}

\section*{Acknowledgements}
Seema Satin is thankful to Ghanshyam Date, Pankaj Joshi and Sukratu Barve for 
helpful discussions.

\end{document}